\documentclass[aps,prl,reprint,twocolumn,showpacs,amsmath,amssymb]{revtex4-1}

\usepackage{graphicx}
\usepackage{amssymb}
\usepackage{amsmath}
\usepackage{dcolumn}
\usepackage{bm}
\usepackage{color}
\usepackage{hyperref}

\begin{document}

\title{Remote Macroscopic Entanglement on a Photonic Crystal Architecture}

\author{H. Flayac}
\email{hugo.flayac@epfl.ch}
\author{M. Minkov}
\author{V. Savona}
\affiliation{Institute of Theoretical Physics, \'{E}cole Polytechnique F\'{e}d\'{e}rale de Lausanne EPFL, CH-1015 Lausanne, Switzerland}

\begin{abstract}
The outstanding progress in nanostructure fabrication and cooling technologies allows what was unthinkable a few decades ago: bringing {single-mode} mechanical vibrations to the quantum regime. {The coupling between photon and phonon excitations is a natural source of nonclassical states of light and mechanical vibrations, and its study within the field of cavity optomechanics is developing lightning-fast. Photonic crystal cavities are highly integrable architectures that have demonstrated the strongest optomechanical coupling to date, and should therefore play a central role for such hybrid quantum state engineering.} In this context, we propose a realistic heralding protocol for the on-chip preparation of remotely entangled mechanical states, relying on the state-of-the-art optomechanical parameters of a silicon-based nanobeam structure. Pulsed sideband excitation of a Stokes process, combined with single photon detection, allows writing a delocalised mechanical Bell state in the system, signatures of which can then be read out in the optical field. A measure of entanglement in this protocol is provided by the visibility of a characteristic quantum interference pattern in the emitted light.
\end{abstract}
\maketitle

\textbf{Introduction.}
Cavity optomechanics \cite{Kippenberg2008,Favero2009,Marquardt2009,Aspelmeyer2014} remarkably allows to transpose the cavity quantum electrodynamics features to vibrational quanta, through radiation pressure. This field of research has demonstrated a spectacular theoretical and experimental development in the past decade and is now envisioned as one of the most promising routes to the production of nonclassical states of a macroscopic degrees of freedom. Recently a number of theoretical investigations have predicted such nonclassical mechanical states to occur in the form of squeezed states \cite{Corbitt2006,Purdy2013,Peano2015}, Fock states \cite{Vanner2013,Galland2014} and photon-phonon \cite{Pirandola2006,Vitali2007,Zhou2011,Akram2012,Ghobadi2014} or purely mechanical entangled states \cite{Pinard2005,Hartmann2008,Ludwig2010,Borkje2011,Ge2013,Akram2013,Tan2013a,Xu2013,Szorkovszky2014,Flayac2014}. Such achievements would not only shine light on the foundations of quantum mechanics and the decoherence process, but could also provide a long-lived information storage or processing platform useful for potential quantum repeaters \cite{Lvovsky2009}. Outstanding experimental results have already been obtained in this direction by Lee and coworkers \cite{Lee2011} who managed, based on the Duan-Lukin-Cirac-Zoller protocol \cite{Duan2001}, to entangle two optical phonon modes stored in remote diamond crystals at room temperature. Meanwhile, current cooling technology \cite{OConnell2010,Chan2011,Riviere2013,Meenehan2014} is paving the way to nonclassical states stored in acoustic phonon modes, which can be pictured as the collective motion of a macroscopic number of atoms in the solid.

The recent prowess in the optimisation of silicon \cite{Chan2011,Safavi-Naeini2014}, indium phosphide \cite{Gavartin2011,Makles2015} or more recently diamond-based \cite{Rath2013,Khanaliloo2015} photonic crystal structures allow to combine high quality factors of the cavity modes with confined mechanical modes, resulting in record values of the optomechanical coupling reaching the MHz range \cite{Chan2011}. Thanks to their small footprint, solid state structures lie among the best candidates for the future of integrated quantum logic elements that could exploit optical and mechanical quanta to code information. In this framework, it is desirable to have access to on-chip operations that involve single quanta of excitation and remotely entangled states \cite{Horodecki2009} of high purity.

In this article we predict the strong potential of optimised silicon nanobeam structures for on-chip entanglement of distant mechanical vibrations. We first design and characterise a promising candidate system based on a state-of-the-art photonic-phononic crystal structure. Then, we demonstrate the possibility to engineer remotely entangled states of mechanical quanta on chip, based on a high efficiency heralding protocol \cite{Flayac2014}. The quantum dynamics is described by means of a master equation treatment that includes the impact of environmental interactions and finite temperature.
\\
\begin{figure*}[ht]
\includegraphics[width=0.8\textwidth,clip]{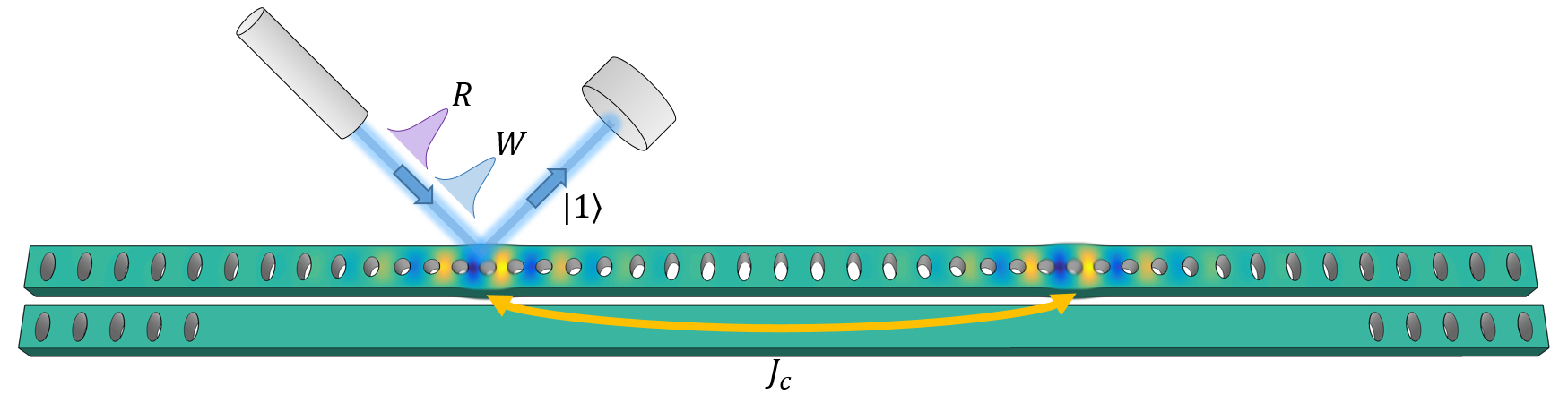}\\
\caption{3D model of the system under study, showing the nanobeam photonic crystal (upper) and the clogged waveguide (lower), together with a sketch of the excitation and detection conditions. The cavity 1 is driven under pulsed excitation and its output is collected by a single photon detector. The yellow arrow highlights the waveguide-induced optical coupling of strength $J_c$ between the 2 cavities. The deformation of the cavities illustrate the slight transverse vibrations induced by the optical field that are computed in Figs.\ref{Fig1}(d) and (e).}
\label{Fig1Large}
\end{figure*}

\textbf{The structure.}
We propose the system shown in Fig.\ref{Fig1Large}, which is based on a pair of silicon nanobeams of width $529$nm and thickness $220$nm, at a distance of $140$nm from each other. The first beam is patterned in order to host two defect cavities based on the optimised design proposed in Refs.\cite{Chan2011, Chan2012}, separated by a distance of $d=$ 6.3 $\mu$m. The lattice constant of the photonic crystal pattern (cf. Ref.\cite{Chan2012} for details) is $a_0 = 436$nm. The second beam hosts a clogged waveguide section that supports Fabry-Perot-like modes and serves to couple the two cavities in an implementation similar to the one of Ref.\cite{Sato2012}. Such a structure could be used to remotely couple cavities that are separated by a distance of hundreds of wavelengths.

\begin{figure}[ht]
\includegraphics[width=0.48\textwidth,clip]{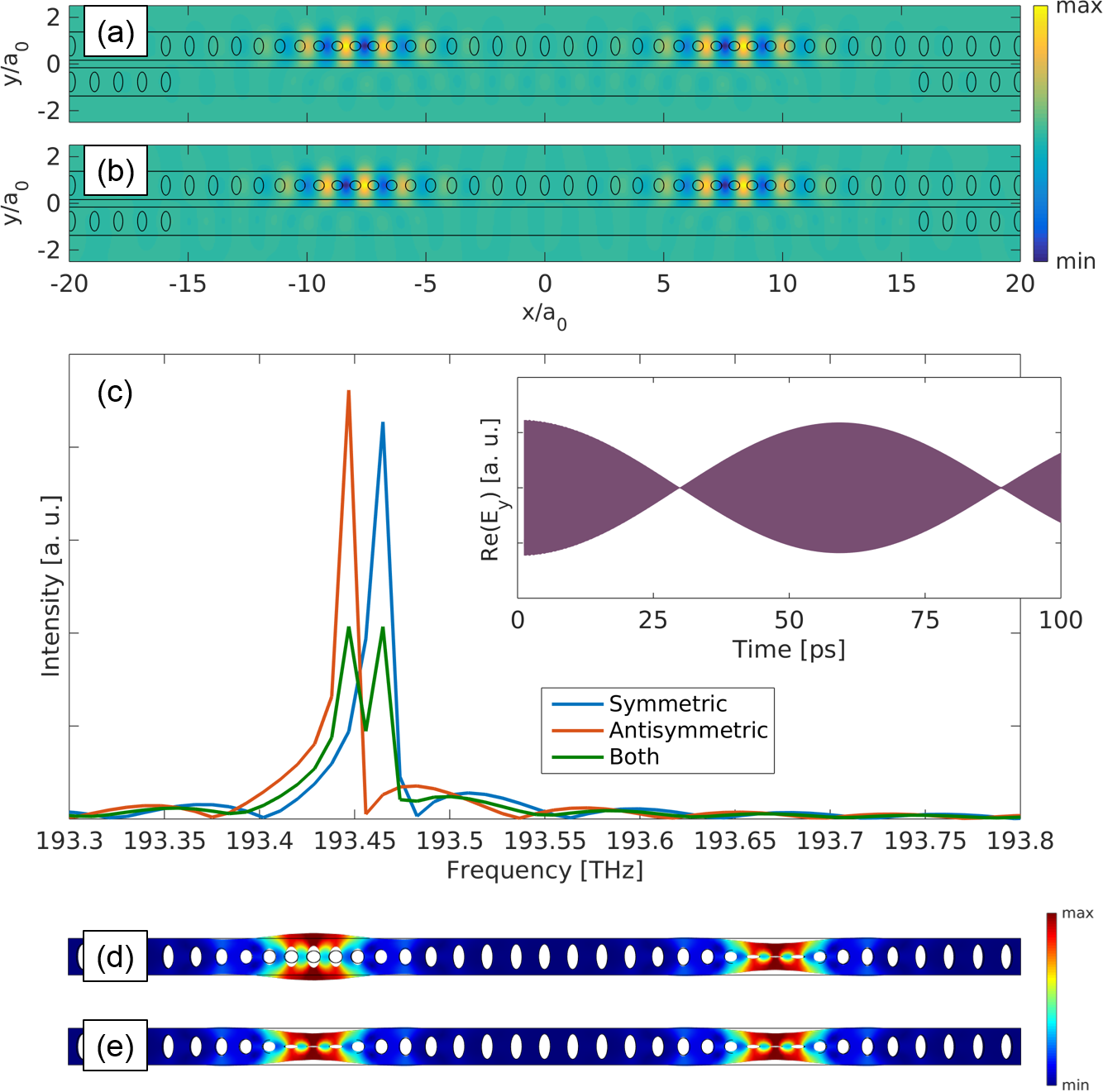}\\
\caption{(a) Anti-symmetric and (b) symmetric normal modes of the electromagnetic field in the coupled cavity system. The colormap expresses the $\rm{Re}(E_y)$ component of the electric field. (c) Simulated optical spectrum after individual or simultaneous excitation of the optical modes. The inset shows the time dependence of the electric field, as recorded at the center of one cavity when both modes are excited. (d) Anti-symmetric and (e) symmetric confined mechanical modes. The colormap expresses the amplitude of the displacement field.}
\label{Fig1}
\end{figure}

To demonstrate the waveguide-mediated optical coupling we model the structure using a commercial-grade simulator based on the finite-difference time-domain (FDTD) method \cite{Lumerical}. By imposing odd, even, or no parity with respect to the $x = 0$ plane, we can selectively excite the anti-symmetric (Fig.\ref{Fig1}(a)), symmetric (Fig.\ref{Fig1}(b)), or both modes of the coupled system. The corresponding simulated spectra are presented in Fig.\ref{Fig1}(c). The modes resonate around the frequency $\omega_c=2\pi\times$193.45 THz. To accurately determine the frequency splitting, we record the electric field in the center of one cavity when both modes are excited, and observe a clear beating pattern, as shown in the inset. From the time interval between the two antinodes we compute $2J_c=2\pi\times16.9$ GHz, which is consistent with the spectrum in Fig.\ref{Fig1}(c). In the absence of the clogged waveguide instead, the simulation results in a splitting of about 4 GHz that would dramatically drop for larger cavity separation, thus highlighting the role of the waveguide in enhancing the coupling between the two cavities. The quality factors of the two modes were computed from the time decay of the field after selective excitation and are $Q_-=1.14\times10^{6}$ and $Q_+=8.0\times10^{5}$, corresponding to linewidths of $\kappa_-=2 \pi\times169$ MHz and $\kappa_+=2 \pi\times243$ MHz for the antisymmetric and symmetric modes, respectively. The linewidths are slightly unbalanced due to the different overlap of the two confined modes with the continuum of radiation modes outside the beams. Even though such high $Q$-s are experimentally achievable \cite{Chan2012}, below we shall conservatively assume twice larger values of $\kappa_-= 2 \pi\times 338$ MHz and $\kappa_+=2 \pi\times486$ MHz in view of the extra losses that would be introduced by the measurement though input and output channels such as tapered fibers \cite{Barclay2005}.

The vibrational eigenmodes of the structure, associated to the two defect cavities, were computed via finite element modeling \cite{Comsol}. The localised transverse breathing modes, as described in Ref.\cite{Chan2011}, are found at a frequency $\Omega_m=2\pi\times5.08$ GHz. As for the optical modes, symmetric and anti-symmetric normal modes of the vibrations at the two cavities arise, which are characterised by a negligible frequency separation of $2J_m = 2\pi\times$8.8 kHz. The displacement fields of these modes are depicted in Fig.\ref{Fig1}(d),(e). The associated distortion of the cavity is oriented mostly along the $y$-direction, producing a record optomechanical coupling of $g = 2\pi\times0.86$ MHz \cite{Chan2012}. We assume the mechanical quality factor $Q_m=6.8\times10^5$ measured in Ref.\cite{Chan2011}, where the structure was coupled to an acoustic radiation shield. As we will see, the readout of the mechanical entangled state requires that the vibrational modes are nondegenerate, in order to introduce a relative phase factor between the two optomechanical coupling processes, that reveals the quantum superposition of the two mechanical modes through an interference pattern in the optical readout \cite{Flayac2014}. Different frequencies for the mechanical modes of the two cavities will naturally arise due to imperfections in the nanofabrication process. We have therefore studied the impact of the structural disorder on the resonant mechanical frequency of one cavity, by introducing a random, Gauss distributed deviation of the holes dimensions and positions with a standard deviation of 1 nm. Over 100 simulated disorder realisations we found a typical 1$\%$ fluctuation of $\Omega_m$ that shall therefore be assumed from now on namely $\Omega_{m,2}=1.01\times\Omega_{m,2}=2\pi\times5.13$ GHz.
\\

\begin{figure}[ht]
\includegraphics[width=0.48\textwidth,clip]{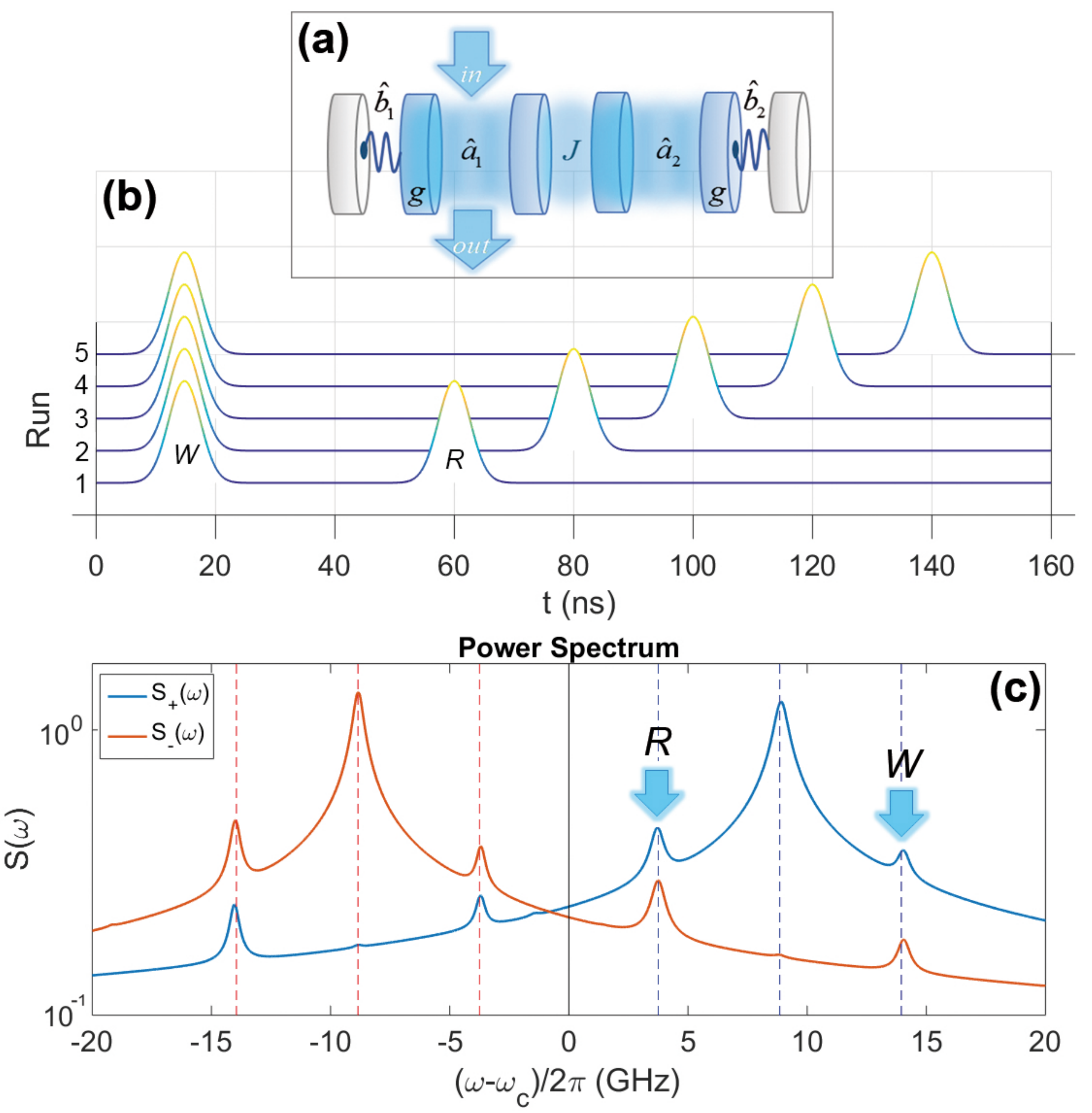}\\
\caption{(a) Schematic representation of the model system. (b) Write and readout pulses sequences for the interference reconstruction. (c) The computed power spectrum of the symmetric and antisymmetric cavity modes, revealing the stokes and antistokes sidebands.}
\label{Fig2}
\end{figure}

\textbf{Quantum model.}
The system can be schematically summarised as in Fig.\ref{Fig2}(a) namely as two coupled, single-mode cavities ($\hat{a}_{1,2}$ operators), each of them optomechanically coupled to distinct mechanical modes ($\hat{b}_{1,2}$ operators). The system Hamiltonian reads
\begin{eqnarray}
{\hat {\cal H}} &=& \sum\limits_{j = 1,2} {\left[ {{\omega _c}\hat a_j^\dag {{\hat a}_j} + {\Omega _j}\hat b_j^\dag {{\hat b}_j} + g\hat a_j^\dag {{\hat a}_j}\left( {\hat b_j^\dag  + {{\hat b}_j}} \right)} \right]}\\
\nonumber &+& {J_c}\left( {\hat a_1^\dag {{\hat a}_2} + \hat a_2^\dag {{\hat a}_1}} \right) + {F}\left( t\right)\left({\hat a_1^\dag {e^{ - i{\omega _L}t}} + {{\hat a}_1}{e^{ + i{\omega _L}t}}} \right).
\end{eqnarray}
We introduce an optical write and readout protocol analogous to the one we recently adopted in the case of a single optical cavity \cite{Flayac2014}. The system is driven first by a classical source of amplitude $F(t)$ and frequency $\omega_L$ on the cavity 1 that stands for our input-output mode (see $W$ in Fig.\ref{Fig2}(c)). It is instructive to rewrite ${\hat {\cal H}}$ on the basis of the symmetric and antisymmetric modes following $\hat a_\pm=(\hat a_1 \pm \hat a_2)/\sqrt{2}$ and $\hat b_\pm=(\hat b_1 \pm \hat b_2)/\sqrt{2}$ which recasts the optomechanical contribution ${\cal{\hat H}}_g = g\sum\nolimits_j {\hat a_j^\dag {{\hat a}_j}( {\hat b_j^\dag  + {{\hat b}_j}} )}$ as
\begin{eqnarray}
\label{Hgpm}
{{\hat {\cal H}}_g} &=& \frac{g}{{\sqrt 2 }}\left( {\hat a_ + ^\dagger {{\hat a}_ + } + \hat a_ - ^\dagger {{\hat a}_ - }} \right)\left( {\hat b_ + ^\dagger  + {{\hat b}_ + }} \right)\\
\nonumber &+& \frac{g}{{\sqrt 2 }}\left( {\hat a_ + ^\dagger {{\hat a}_ - } + \hat a_ - ^\dagger {{\hat a}_ + }} \right)\left( {\hat b_ - ^\dagger  + {{\hat b}_ - }} \right).
\end{eqnarray}
It describes two cavity modes of resonance $\omega_\pm=\omega_c\pm J_c$ (i) optomechanically coupled to the same vibrational mode $\hat b_+$ of frequency $\Omega_+=(\Omega_1+\Omega_2)/2\simeq\Omega_{1,2}$ (first term) and (ii) coupled together through the absorbtion or emission of a phonon of the mode $\hat b_-$ (second term). The latter process, that would allow an exchange of quanta between the $\hat a_+$ and $\hat a_-$ modes, is off-resonant in the conditions we consider where $\omega_+-\omega_-\simeq3\Omega_{1,2}$ and can be discarded. The cavity spectrum shown in Fig.\ref{Fig2}(c) consists of two main resonances at $\omega_c\pm J$ each surrounded by sidebands formed by the Raman interaction with the $\hat{b}_+$ mode (dashed lines). Note that (ii) makes the $\hat a_+$ sidebands visible in the $\hat a_-$ spectrum and vice versa.

If the cavity 1 is driven on the resolved $\hat{a}_+$ antistokes sideband, i.e. setting $\omega_L=\omega_c + J + \Omega_+$ (see $W$ arrow in Fig.\ref{Fig2}(c)), the first term of Eq.(\ref{Hgpm}) then produces the two mode squeezing \cite{Aspelmeyer2014}
\begin{equation}\label{Hglin}
{{\hat {\cal H}}_g} \approx \frac{{g}}{{\sqrt 2 }}\left({\hat a_ + ^\dagger } \hat b_ + ^\dagger  + {{{\hat a}_ + }}{{\hat b}_ + }\right),
\end{equation}
The $\hat a_-$ contribution is negligible since it is shifted by $-(2J+\Omega_+)$ from the pump frequency which makes it strongly nonresonant. The interaction (\ref{Hglin}) therefore creates a pair correlation between the symmetric cavity fluctuations and mechanical modes, resulting in a squeezed state of the mode $\hat{b}_+$. Under continuous wave driving, one could therefore achieve a steady state hybrid continuous variable entanglement between the $\hat{a}_+$ and $\hat{b}_+$ modes \cite{Vitali2007}. However, the goal here is to produce a mechanical excitation in a linear superposition of two Fock states, each corresponding to one quantum of vibration in one cavity. This requires combining pulsed excitation with postselection. Indeed, after the write pulse, the detection of a single photon emitted by any of the two cavities heralds the presence of a single phonon \cite{Galland2014,Flayac2014} stored in the mechanical mode $\hat b_+$, i.e. shared between the remote modes $\hat b_1$ and $\hat b_2$.

Assuming zero temperature, the system lies initially in its vacuum state $\left| {{\psi _0}} \right\rangle = {\left| 0 \right\rangle _{{c_1}}} \otimes {\left| {00} \right\rangle _m} \otimes {\left| 0 \right\rangle _{{c_2}}}$, where the indices $c_j$ and $m$ refer to the cavities and mechanical states respectively. The write laser tuned to the Stokes process of the $\hat a_+$ mode ideally results in the state
\begin{eqnarray}
\nonumber{{\hat {\cal H}}_g}\left| {{\psi _0}} \right\rangle
          &=& c_{00}{\left| 0 \right\rangle _{{c_1}}} \otimes {\left| {00} \right\rangle _m} \otimes {\left| 0 \right\rangle _{{c_2}}}\\
          \label{psi1}
          &+& c_{10}{\left| 1 \right\rangle _{{c_1}}} \otimes \left( {{{\left| {10} \right\rangle }_m} + {e^{i\varphi }}{{\left| {01} \right\rangle }_m}} \right) \otimes {\left| 0 \right\rangle _{{c_2}}} \\
\nonumber &+& c_{01}{\left| 0 \right\rangle _{{c_1}}} \otimes \left( {{{\left| {10} \right\rangle }_m} + {e^{i\varphi }}{{\left| {01} \right\rangle }_m}} \right) \otimes {\left| 1 \right\rangle _{{c_2}}} + \ldots\,
\end{eqnarray}
Here $\varphi=\varphi_0+(\Omega_2-\Omega_1)t$ where $\varphi_0$ is a constant phase shift depending on the details of the excitation scheme, and the coefficients $c_{ij}$ are determined by the driving field. In particular one expects that, in the limit of low driving field, the components of this state corresponding to two- and more photons are negligible. A single photon detection from the cavity 1 retains only the second term of Eq.(\ref{psi1}) and the state of the system, after detection, reduces to
\begin{eqnarray}
\label{psiW}
  {\left| {{\psi_{W}}} \right\rangle} = c_{10}^\prime{\left| 0 \right\rangle _{{c_1}}} \otimes \left( {{{\left| {10} \right\rangle }_m} + {e^{i\varphi^{\prime} }}{{\left| {01} \right\rangle }_m}} \right) \otimes {\left| 0 \right\rangle _{{c_2}}}
\end{eqnarray}
which is nothing but a  maximally entangled mechanical Bell state ${\left| \psi_{\cal{H}}  \right\rangle _m} = \left( {{{\left| {10} \right\rangle }_m} + \exp(i\varphi^{\prime}){{\left| {01} \right\rangle }_m}} \right)/\sqrt 2$ when the cavity modes are traced out.

The nonclassical character of the mechanical quantum state can only be indirectly assessed through the optical field. For this, the information stored in the phonon state has to be transferred back to the cavity field. This is achieved thanks to the anti-Stokes process induced by a second pulse of frequency $\omega_L=\omega_c + J - \Omega_+$ onto the cavity 1 (see $R$ arrow in Fig.\ref{Fig2}(c)). Such excitation condition is resonant with the beam splitter interaction \cite{Aspelmeyer2014}
\begin{equation}
\label{Hgread}
{{\hat {\cal H}}_g} \approx \frac{{g}}{{\sqrt 2 }} \left({\hat a_ + ^\dagger } \hat b_ +  + {{{\hat a}_ + }}{{\hat b}_ + ^\dagger }\right),
\end{equation}
that swaps the $\hat b_+$ phonon with an $\hat a_ +$ photon. When (\ref{Hgread}) is applied to the written state one obtains
\begin{eqnarray}
\nonumber \left| {{\psi _{R}}} \right\rangle  &=& c_{00}^{\prime \prime }{\left| 0 \right\rangle _{{c_1}}} \otimes \left( {1 + {e^{i\varphi^{\prime\prime} }}} \right){\left| {11} \right\rangle _m} \otimes {\left| 0 \right\rangle _{{c_2}}} \\
          &+& c_{10}^{\prime \prime }{\left| 1 \right\rangle _{{c_1}}} \otimes \left( {1 + {e^{i\varphi^{\prime\prime} }}} \right){\left| {00} \right\rangle _m} \otimes {\left| 0 \right\rangle _{{c_2}}} \\
\nonumber &+& c_{01}^{\prime \prime }{\left| 0 \right\rangle _{{c_1}}} \otimes \left( {1 + {e^{i\varphi^{\prime\prime} }}} \right){\left| {00} \right\rangle _m} \otimes {\left| 1 \right\rangle _{{c_2}}} +  \ldots
\end{eqnarray}
The corresponding cavity 1 intensity ${\cal{I}}_{1} = \left\langle {{\psi _{R}}} \right|\hat a_{1}^\dag {{\hat a}_{1}}\left| {{\psi _{R}}} \right\rangle = |c_{10}^{\prime\prime}|^2 [1 + \cos \left( \varphi^{\prime\prime}  \right)]$, where $\varphi^{\prime\prime}=\varphi_0 + \Delta\Omega_m t^{\prime\prime}$ depends on the delay from the write procedure and presents a full contrast interference pattern providing means to reveal the entangled nature of the mechanical state. Note that one could equally detect the cavity 2 field and obtain the same result. The expected interferences can be reconstructed by integrating the emission received by a single photon detector over several runs where the delay between the write and readout pulse is gradually increased as shown in Fig.\ref{Fig2}(b) and detailed below.

The simple argument based on pure states that was presented above can be supported by a numerical study of the full system dynamics, that we carried out using the theoretical tools to model open quantum systems \cite{Gardiner2010}. In this analysis, in order to be able to account for the sizeable classical field component induced by the driving laser, we separated the field into the sum of its classical and quantum-fluctuation components according to $\hat{a}_{\pm}=\alpha_{\pm}+\delta\hat{a}_{\pm}$ and $\hat{b}_{\pm}=\beta_{\pm}+\delta\hat{b}_{\pm}$. The classical evolution is governed by the following set of equations
\begin{eqnarray}
\nonumber  i{{\dot \alpha }_ \pm } &=& \left( {\pm J + g\sqrt 2 \operatorname{Re} \left( {{\beta _ \pm }} \right) - i\frac{{{\kappa _ \pm }}}{2}} \right){\alpha _ \pm }\\
&+& g\sqrt 2 \operatorname{Re} \left( {{\beta _ \mp }} \right){\alpha _ \mp } + \frac{F\left( t \right)}{{\sqrt 2 }} \\
\nonumber i{{\dot \beta }_ \pm } &=& \left( {{\Omega _ + } - i\frac{{{\gamma _ \pm }}}{2}} \right){\beta _ \pm } + {\Omega _ - }{\beta _ \mp }\\
&+& \frac{g}{{\sqrt 2 }}\left( {\alpha _ + ^*{\alpha _ \pm } + \alpha _ - ^*{\alpha _ \mp }} \right)
\end{eqnarray}
that are written here in the frame rotating at the cavity frequency $\omega_c$ and on the basis of symmetric and antisymmetric modes $\alpha_{\pm}$ and $\beta_{\pm}$ to account for their unbalanced losses $\kappa_\pm$. We however consider equal mechanical losses $\gamma_{\pm}=2\pi\times3.75$ kHz which is valid given the large time scales involved. The dynamics of quantum fluctuations is treated via a master equation for the density matrix in the displaced reference frame set by the classical field \cite{Nunnenkamp2011}
\begin{eqnarray}
\frac{{d\hat \rho }}{{dt}} &=&  - i\left[ {{\cal{\hat{H}}}_d,\hat \rho } \right] - \frac{1}{2}\sum\limits_{j =  \pm } { {{\kappa _ j }{\cal{\hat{D}}}[{{\delta\hat a}_ j }]\hat \rho} }\\
\nonumber &-& \frac{1}{2}\sum\limits_{j =  \pm } {\left[ {{\gamma _ j }\left( {{{\bar n}_{th}} + 1} \right){\cal{\hat{D}}}[{{\delta\hat b}_ j }]\hat \rho  + {\gamma _ j }{{\bar n}_{th}}{\cal{\hat{D}}}[\delta\hat b_ j ^\dag ]\hat \rho } \right]}
\end{eqnarray}
where ${\bar n}_{th}$ is the mean thermal phonon occupation and ${\cal{\hat{D}}}\left[ {\hat o} \right]\hat \rho  = {{\hat o}^\dag }\hat o\hat \rho  + \hat \rho {{\hat o}^\dag }\hat o - 2\hat o\hat \rho {{\hat o}^\dag }$ describe the dissipations in the environment for each mode.
\\

\begin{figure}[ht]
\includegraphics[width=0.48\textwidth,clip]{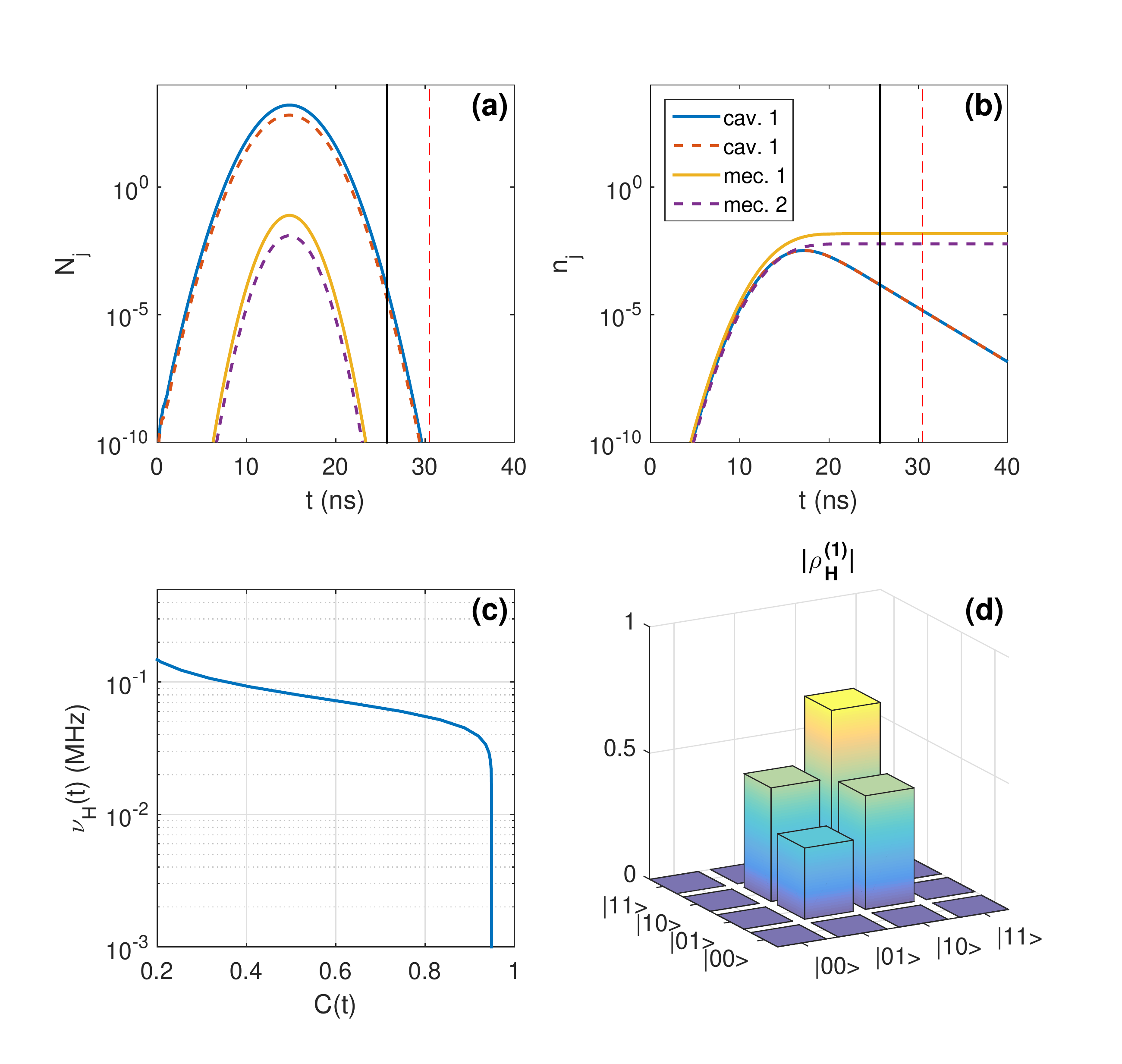}\\
\caption{Write procedure: Log scale (a) classical and (b) fluctuation occupations versus time. Blue and dashed-red lines: cavity mode 1 and 2; Yellow and dashed purple lines: mechanical mode 1 and 2. The vertical black line marks the time $t_q$, starting from which quantum fluctuations become dominant. (c) The computed heralding rate $\nu_{\cal{H}}(t)$ versus the concurrence ${\cal{C}}(t)$. (d) Absolute value of the heralded mechanical density matrix $\hat \rho_{\cal{H}}^{(1)}$ at the ${\cal{C}}_{\rm{max}}$ time (vertical red dashed line in (a) and (b)) conditioned on the detection of a single photon out of  cavity 1.}
\label{Fig3}
\end{figure}

\textbf{Results and discussion.}
The write procedure is conducted with a Gaussian excitation at a time $t_{{W}}=30$ ns, of amplitude ${\cal{A}}_{{W}} = 10^{3}\kappa_-$, standard deviation $\sigma_t = 3.85$ ns and frequency $\omega_L = \omega_c + J + \Omega_+$ on the cavity 1. Note that one could equally excite the antisymmetric mode at $\omega_L = \omega_c - J + \Omega_+$ as prescribed by \eqref{Hgpm}. The driving pulse results in a large classical field amplitude for the cavity mode. This acts as a source for the mechanical field and produces a sizeable phonon occupation in both mechanical modes, thanks to the optical coupling between the two cavities. In this way, the normal modes $\hat{a}_+$ and $\hat{b}_+$ are excited without needing to drive the field in the second cavity. Figures \ref{Fig3}(a) and (b) show respectively the classical and quantum fluctuation occupations. We observe that the classical components vanish along with the input pulse. Starting from a time $t_q\simeq25$ ns (vertical black line), quantum fluctuations dominate over the classical components and decay according to the respective lifetimes. The quantity $t_q$ defines a time horizon past which the system evolves freely and can display quantum correlations. Importantly, a compromise must be made on the pulse duration, which should be set (i) long enough to accurately excite the sideband without leaking too much on the main cavity frequency and (ii) short enough to allow for a sufficiently large occupation of the quantum fluctuations at time $t_q$.

The heralding at a time $t_{\cal{H}}>t_q$ is modeled via the projection of the density matrix onto the single photon subspace \cite{Galland2014,Flayac2014} of cavity 1
\begin{eqnarray}
\label{rhoH}
{{\hat \rho }_{\cal{H}}}^{(1)} &=& \frac{{\hat{\cal{P}}_{1}\hat \rho \left( t_{\cal{H}} \right)\hat{\cal{P}}_{1}^{-1}}}{{{\rm{Tr}}\left[ {\hat {\cal{P}}_{1}\hat \rho \left( t_{\cal{H}} \right)} \right]}},\label{rhop}\\
{{\hat {\cal{P}}}_{1}} &=& {\left| 1 \right\rangle}_{c_1}\langle 1{|}_{c_1} \otimes {\mathbb{\hat{I}}_{{m_1}}} \otimes {\mathbb{\hat{I}}_{{m_2}}} \otimes {\mathbb{\hat{I}}_{c_2}}
\end{eqnarray}
We consider in Eq.(\ref{rhoH}) the full density matrix of the system, obtained by including the classical contributions as coherent states $\left| {{\alpha _ \pm }} \right\rangle$ and $\left| {{\beta _ \pm }} \right\rangle$ and applying a multimode displacement operator ${\hat {\cal{D}}}(\alpha_\pm,\beta_\pm)$ to $\hat\rho$. The heralding efficiency -- therefore the degree of mechanical entanglement -- is computed through the \emph{concurrence} ${\cal{C}}$ \cite{Wootters1998} of the reduced density matrix $\hat\rho_m=\rm{Tr}_{\hat{a}_1,\hat{a}_2}({\hat\rho_{\cal{H}}})$ restrained to the 0 and 1 phonon subspaces. It displays a large maximum value of ${\cal{C}}_{\rm{max}}=0.95$ from $t_{\cal{H}}=30$ ns (see vertical dashed lines in Fig.\ref{Fig3}(a),(b)).  Corresponding to this value, a mechanical Bell state storing delocalised phonon is formed, as revealed by the $|\hat\rho_{\cal{H}}^{(1)}|$ visual shown in Fig.\ref{Fig3}(d). The imbalance in the diagonal elements is a consequence of having driven a single cavity only. We have computed a fidelity ${\cal{F}}=0.97$ to the $(\left| {10} \right\rangle  + \left| {01} \right\rangle)/\sqrt{2}$ Bell state. Obviously, the heralding time would be random in an experiment and the detector should target the $t>t_q$ time window where ${\cal{C}}\sim{\cal{C}}_{\rm{max}}$, while avoiding to monitor earlier times, for the procedure to succeed. The entangled state lives as long as the mechanical lifetime $\tau_m=2 \pi/\gamma_{1,2}$. Given the very high quality factor characterising nanobeam cavities, combined to the GHz-range of the frequency of the mechanical modes, one could experimentally reach $\tau_m = 277$ ms \cite{Chan2012} which is promising for quantum repeater technology \cite{Lvovsky2009}. When aiming at a concurrence ${\cal{C}}>0.9$, the largest heralding rate achieved in our simulations was $\nu_{\cal{H}}=\kappa_{1,2}n_{1,2}\sim3\times10^{-2}$ MHz, where $\kappa_{1,2}=(\kappa_+ +\kappa_-)/2$, as shown in the figure \ref{Fig3}(c). Finally we point out that, in addition to a large value of the simulated concurrence, in order to assess our target entangled state it is required that the two- and higher- photon terms in the full mechanical density matrix be negligible as it is the case here. This is simply achieved by controlling the amplitude of the driving field.

\begin{figure}[ht]
\includegraphics[width=0.48\textwidth,clip]{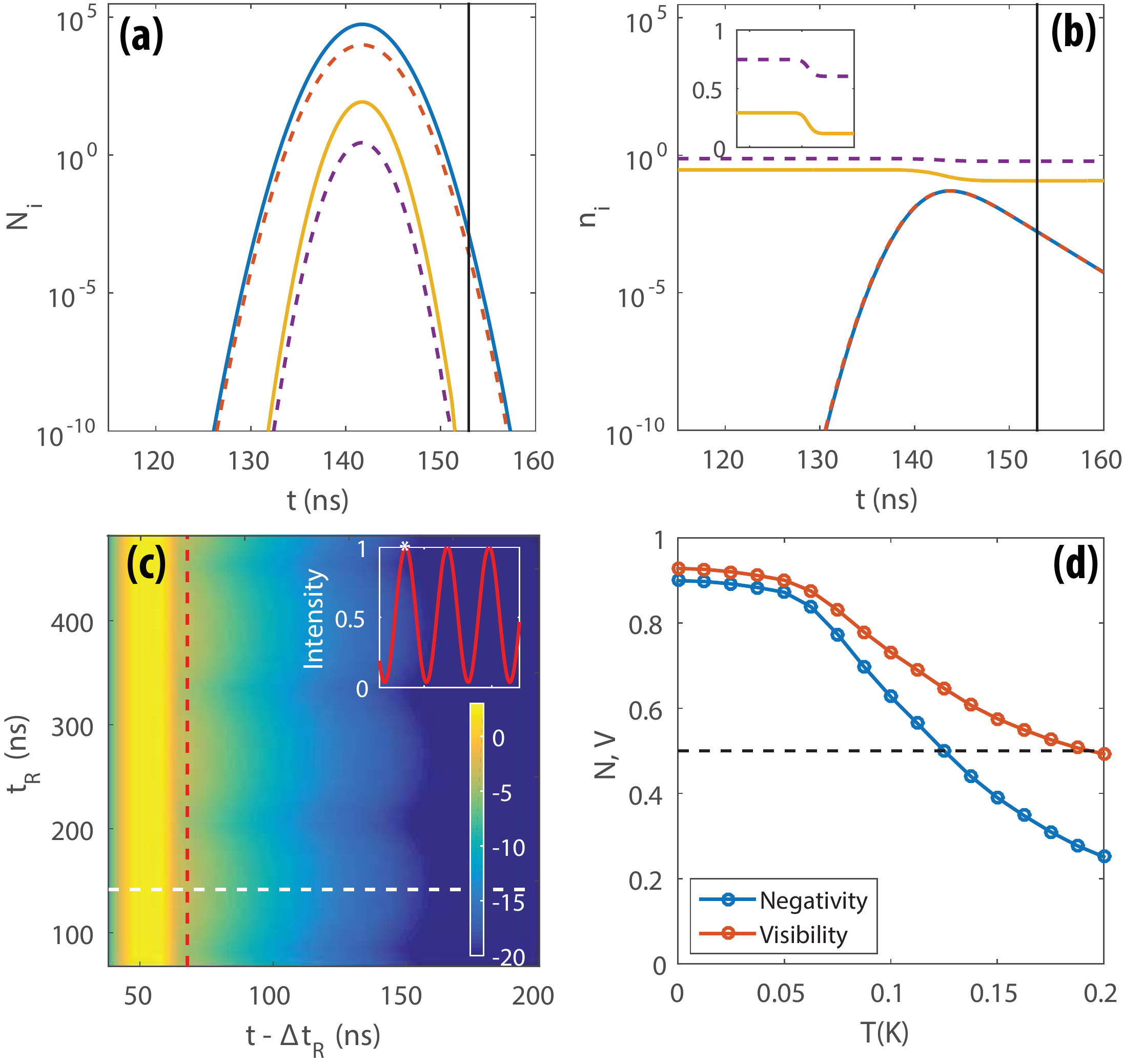}\\
\caption{Readout procedure: Logarithmic scale (a) classical and (b) fluctuation contributions to the state occupations versus time, as computed for a given realisation of the readout procedure. The inset of panel (b) shows the mechanical occupation in linear scale. (c) The field intensity in cavity 1, averaged over several runs, as a function of the readout $t_R$ and shifted evolution time $t-\Delta t_R$ where $\Delta t_R = t_R^{(j)}-t_R^{(0)}$ for the run number $j$. The inset shows a (normalised) cut along the vertical red dashed line, from which the visibility ${\cal{V}}$ can be inferred. The horizontal white dashed line corresponds to the (a) and (b) evolutions where the intensity is at a maximum. (d) Simulated temperature dependence of the negativity and the fringe visibility.}
\label{Fig4}
\end{figure}

The readout step involves a second input pulse having the same width as the write pulse, and amplitude ${\cal{A}}_{{R}} = 5\times10^{3}\kappa_-$ delayed by $\Delta T=t_{{R}}-t_{{W}}$. The central frequency of the pulse is tuned to the anti-Stokes $\hat a_+$ sideband $\omega_L = \omega_c + J - \Omega_+$, and aims at transferring the information about the phonon mode back to the optical field. As discussed previously, the nondegeneracy of the mechanical modes $\Delta\Omega = \Omega_2-\Omega_1 \simeq$ 50 MHz, resulting from the sample imperfections, introduces a relative phase factor $\varphi = \varphi_0 + \Delta\Omega t$ between the modes which is associated with an interference pattern in the emitted light. The latter is revealed by varying the write and readout pulse delay $\Delta T$ and integrating the corresponding signals (see Fig.\ref{Fig2}(b)). Figures \ref{Fig4}(a),(b) show the classical and fluctuations contributions, respectively, to the occupations in a realisation where the cavity intensity reaches a maximum (dashed white line in panel(c)). Once again, quantum fluctuations dominate after the driving pulse has decayed (vertical black line). Starting from this time, the quantum interference should be recorded. From the panel (b) inset one can see the $\hat b_+$ mode being emptied by the optical field as revealed in the $\hat b_1$ and $\hat b_2$ simultaneous population decay.

Varying gradually the delay $\Delta T$ between the write and readout pulses and specifically $t_R$ reveals the predicted interference pattern of period ${\cal{T}}=2\pi/\Delta\Omega_m\simeq125$ ns, as shown in Fig.\ref{Fig4}(c). The field in cavity 2 obviously displays a similar behavior (not shown). Note that to obtain a vertical pattern in the plot (instead of an oblique one), the readout pulses have been shifted so that the $x$-axis shows the the shifted evolution time $t - \Delta t_{R}$ where $\Delta t_{R}=t_{R}^{(j)}-t_{R}^{(0)}$ and $j$ is the run index. This corresponds to what would be obtained varying the phase $\varphi_0$ of the write pulse instead of $\Delta T$. A normalised cut along the vertical axis (dashed red line), taken at fixed delay from $t_{{R}}$, is plotted in the inset. It shows a fringe visibility as large as ${\cal{V}}\simeq0.92$. The interference carries a signature of the relative phase between the two confined phonon modes. This can be related to the amount of phonon entanglement. In particular~\cite{Flayac2014}, a fully separable state would result in a visibility ${\cal{V}}_c=0.5$. This value sets a lower bound, below which the entanglement may still be present in the system, but cannot be inferred from the interference pattern. The average cavity occupation at this value of $t_{{R}}$ amounts to $n_{1}\simeq5\times10^{-4}$, which corresponds to a photoemission rate of $\nu=0.2$ MHz. This value is a consequence of the large driving that temporarily displaces the system to a coherent state \cite{Bruno2013} until its quantum nature is restored.

At zero temperature we obtained a Bell state with high fidelity, which could be characterised by the concurrence, if restricting the analysis to the subspace with 0- and 1-phonon states. However, at finite temperature a non-negligible thermal occupation may result in a sizable contribution from states with $n>1$ phonons. In this case, an entanglement measure appropriate to the larger Hilbert space is needed. The concurrence was recently extended to 4-dimensional spaces \cite{Audenaert2001}. Its computation however is a demanding numerical task for $n\ge4$ \cite{Rungta2001,Rungta2003} and is therefore not appropriate in the general case we want to consider here. The Bell states naturally suggest to deploy an entanglement monotone derived from positive partial transpose criterion, namely the \emph{negativity} \cite{Vidal2002} ${\cal{N}}={|| {\hat \rho _m^{{\Gamma _{1,2}}}} ||_1} - 1$, where ${\hat \rho _m^{{\Gamma _{1,2}}}}$ is the partial transpose of $\hat \rho_m$ and ${|| {\hat \rho _m^{{\Gamma _{1,2}}}} ||_1}$ its trace norm. The above argument using quantum states can be extended to mixed states, in order to estimate a lower bound on the visibility resulting from an entangled state, assuming in particular that the mechanical modes initially lie in the product of 2 thermal states. In the supplementary information \cite{Supplemental} we analytically derive the Heralded mechanical density matrix at finite temperature, and recover the lower bound of ${\cal{V}}\simeq0.5$ to the fringe visibility produced by an entangled state.

Besides, the impact of state-of-the-art cooling temperature $T$ \cite{Meenehan2014} on our protocol was computed by direct master equation simulations, and the results are presented in Fig.\ref{Fig4}(d), where ${\cal{N}}$ and ${\cal{V}}$ versus $T$ is plotted. We see, as expected, a clear correlation between ${\cal{N}}$ and ${\cal{V}}$. They both start decreasing from $T\simeq0.05$K. Entanglement survives in the temperature range we considered, and the visibility reaches the lower bound ${\cal{V}}_c$ at $T_c\simeq0.2$K. Below this temperature, our protocol can't assess the presence of entanglement, and a full quantum tomography of the state would be instead required. Finally the impact of the pure dephasing imposed by other mechanical resonances was analysed in Ref.\cite{Flayac2014} and was shown to be negligible in the photonic crystal parameter range.
\\

\textbf{Conclusion.}
We have proposed a realistic photonic crystal architecture based on coupled nanobeams and computed its optomechanical properties. We have shown that this structure should allow to achieve on-chip remote entanglement between the localised acoustic phonon modes, mediated by the optical field, with no need for an additional beam splitter as in typical entanglement protocols \cite{Pirandola2006,Lee2011}. The protocol that we propose achieves high heralding rates and is still effective at typical cooling temperature for Silicon-based systems. The characteristic long mechanical lifetime makes the present system very promising for demonstrating macroscopic entanglement and for the future of hybrid quantum logic circuits.
\\

\textbf{Acknowledgments.}
We thank A. Feofanov for discussions on the experimental feasibility.

\bibliography{Bibliography}

\end{document}